# Enhanced visible light absorption in ZnO/GaN heterostructured nanofilms

Yang Zhang,* Zhi-Feng Wu, Peng-Fei Gao, Dang-Qi Fang and Sheng-Li Zhang*

*Ministry of Education Key Laboratory for Nonequilibrium Synthesis and Modulation of Condensed Matter, Department of Applied Physics, School of Science, Xi'an Jiaotong University, Xi'an 710049, China.*

*Corresponding author: yzhang520@mail.xjtu.edu.cn, zhangsl@mail.xjtu.edu.cn

**Graphical Abstract:**

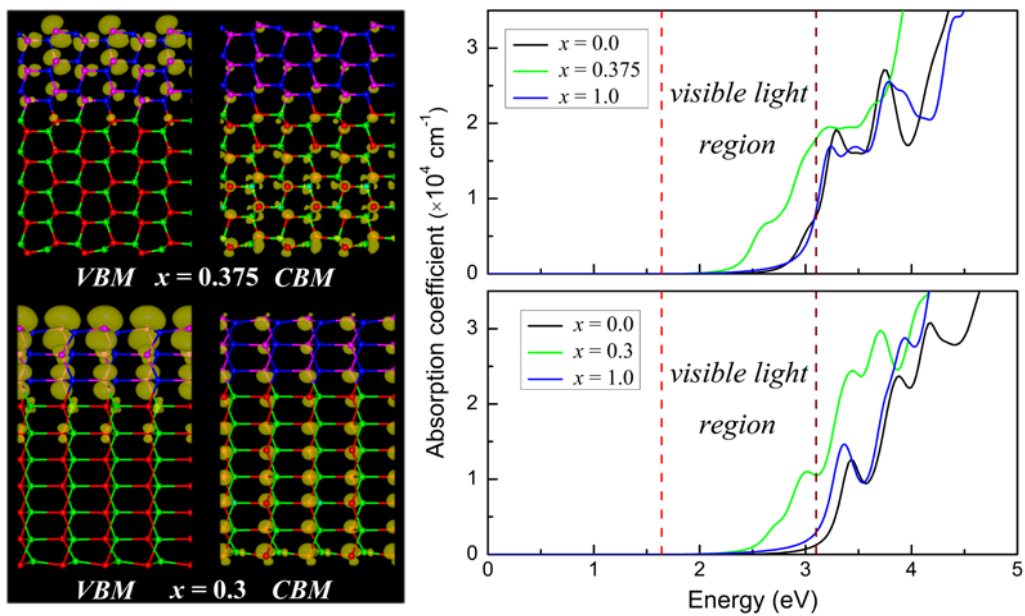

**Highlights:**

1. ZnO/GaN heterostructured nanofilms exhibit flexibly tunable band gaps.

2. ZnO/GaN heterostructured nanofilms show remarkable spatial separation of carriers.

3. Visible light absorption is observed in ZnO/GaN heterostructured nanofilms.

**ABSTRACT:**

First-principles calculations have been employed to investigate structural stability and electronic properties of non-polar $(1\bar{1}00)$ and $(11\bar{2}0)$ ZnO/GaN heterostructured nanofilms. The effects of nanofilm thickness and GaN ratio are considered. It has been found that all studied heterostructured nanofilms are less stable than the corresponding pure ZnO film but more stable than pure GaN one, exhibiting a much thicker film with better stability. Electronic band structures display that both two types of heterostructured nanofilms are semiconductors with their band gaps strongly depending on the GaN ratios as well as the thicknesses. Of particular interest is that the band gaps decrease firstly, and then increase with the increasing GaN ratio, showing flexibly tunable band gaps that cover a wide range of the solar spectrum. Furthermore, spatial charge distribution to the valence band maximum and the conduction band minimum has been studied. By calculating the complex dielectric function, the properties of optical absorption has been explored to exploit their potential application in the solar energy harvesting.

## 1. Introduction

Energy crisis and environmental pollution are two of the most serious problems that are facing to the human society. To solve these problems, utilizing cheap solar energy to produce clean hydrogen energy is considered to be one of the most effective ways. Since the discovery of water splitting into $H_2$ and $O_2$ using $TiO_2$ electrodes under ultraviolet (UV) illumination in 1972 [1], photocatalytic decomposition of water has attracted worldwide attentions [2-4]. Because the UV light in the solar spectrum only accounts for ~ 5 %, while the visible light takes over ~ 50 %, designing of superior photocatalyst driven by the visible light becomes urgent and significant for high efficient utilization of the solar energy.

Owing to the flexibly tunable band gaps that cover a wide range of the solar spectrum, wurtzite gallium zinc oxynitride $(Ga_{1-x}Zn_x)(N_{1-x}O_x)$ has been demonstrated to be a promising photocatalyst that can accomplish overall water splitting under the visible light irradiation [5-12]. The apparent quantum efficiency for water splitting can be as high as 5.9 % in the 420-440 nm wavelength range [5], which cannot be achieved with the presence of sacrificial regents and more than two times higher than that of the previously reported catalyst [6,7]. Unfortunately, this hybrid system is solid solution composed of ZnO and GaN components within single-crystalline wurtzite structure [8-12]. And thus, the distributions of valence band maximum (VBM) and conduction band minimum (CBM) (or photo-generated electrons and holes) mix together in some certain spaces, which results in a low efficient utilization of the solar energy. In addition, numerous studies have suggested that the wurtzite ZnO and GaN could form various heterojunction interfaces with remarkable separated distribution of electrons and holes [13-18]. Due to the effects of type-II band offset [13,14], the ZnO/GaN heterostructured semiconductors present narrow band gaps in the visible-light region and can serve as potential photocatalysts [16,17]. Accordingly, these basic properties, such as suitable band gaps, spatial separation of electron-hole pairs, and visible light absorption are significant for a photocatalyst.

In our previous study [18], we have found that ZnO/GaN heterostructured nanowires exhibit good structural stability and flexibly tunable band structures by their components, suggesting potential applications in solar energy harvesting. Due to the smaller surface to volume ratio, two-dimensional nanofilms are much more stable than one-dimensional nanowires, and could be used in some more extreme environments. Besides, GaN has often been used as the substrate for the growth of ZnO in experiment because of the same wurtzite crystal structure and similar lattice parameters [19-22], which implies that the ZnO/GaN heterostructured nanofilms could be synthesized possibly. Naturally, some questions would be raised: could ZnO/GaN heterostructured nanofilms exhibit semiconducting characteristics with suitable band gaps that are in the range of visible light spectrum? Could they hold great promise for the solar energy utilization? However, these problems are still unclear.

In this study, density functional theory (DFT) calculations have been employed to investigate structural stabilities, electronic properties, and optical absorption of $(1\bar{1}00)$ and $(11\bar{2}0)$ ZnO/GaN heterostructured nanofilms. The influences of GaN ratio and thickness have been explored. It has been found that all heterostructured nanofilms exhibit good structural stabilities and semiconductor characteristics with their band gaps tuned flexibly by the GaN ratio and the thickness. Furthermore, these heterostructured nanofilms exhibit remarkable spatial separation of electrons and holes, and optical absorption in visible light range.

## 2. Computational details

Our study has been performed by employing first-principles calculations based on the spin-polarized DFT within the projector augmented wave method [23,24], as implemented in Vienna *ab initio* simulation package (VASP) [25,26]. The generalized gradient approximation (GGA) and the functional of Perdew-Burke-Ernzerhof (PBE) are used to describe the exchange-correlation interactions [27,28]. The cut-off of plane-wave energy and the convergence of total energy are 450 eV and $10^{-5}$ eV, respectively. For approximating the brillouin zone integrations, $7 \times 7 \times 1$ *k*-point meshes with Gamma centered grid are adopted for ZnO/GaN heterostructured nanofilms. All the nanofilms are modelled in a large rectangular supercell with their surfaces paralleled to the *x-y* plane. Because of the periodic boundary conditions, a vacuum region of at least 10 Å along the *z*-axis (perpendicular to the surface) is applied for eliminating the interactions of neighbor nanofilms. Structural optimizations are performed by computing the Hellmann-Feynman forces using conjugate gradient algorithm within a force convergence of 0.05 eV/Å [29]. For the band gap calculations, the screened hybrid density functional of Heyd-Scuseria-Ernzerhof with 2006 parameterization (HSE06) are employed [30,31], which has been well known as an effective measurement for the band gap of semiconductor [20,32,33].

Since electronic properties of ZnO/GaN heterostructures are strongly dependent on their hetero-interfaces [13-17], two types of non-polar ZnO/GaN heterostructured nanofilms with wurtzite structures, that is, $(1\bar{1}00)$ and $(11\bar{2}0)$ have been considered and shown in Fig. 1. The supercells of $(1\bar{1}00)$ and $(11\bar{2}0)$ heterostructured nanofilms consist of one primitive cell with thicknesses of 8 and 10 layers, and total atoms of 32 and 40, respectively. For comparison with heterostructured nanofilms, pure ZnO and GaN nanofilms with the same thicknesses (i.e., layer numbers) have also been explored. In addition, the effects of GaN ratio $x$ on the structural stabilities and electronic properties of both types of heterostructured nanofilms have been discussed, where the GaN ratio $x$ is defined as $x = n/(m+n)$ ($m$ and $n$ are the numbers of Zn-O and Ga-N pairs in the nanofilms).

## 3. Results and discussion

Structures and electronic properties of wurtzite ZnO and GaN bulks are investigated firstly. The results of optimized lattice constants of $a$ = 3.286 Å, $c$ = 5.312 Å and $u$ = 0.378 ($a$ = 3.248 Å, $c$ = 5.283 Å, and $u$ = 0.377) for the ZnO (GaN) bulk are consistent with the previous theoretical and experimental studies [19-21]. These structural parameters are almost the same along the *a* or *c* axis of the wurtzite ZnO and GaN bulks, exhibiting good lattice matching. Besides, the electronic band structures calculated by using the HSE06 method suggest that both the ZnO and GaN bulks are wide direct band gap semiconductors with the band gaps of 3.190 and 3.409 eV, respectively, being close to the experimental values [34,35].

*3.1 Structural stability*

After structural optimization, both pure ZnO and GaN $(1\bar{1}00)$ and $(11\bar{2}0)$ nanofilms with given 8 and 10 layers exhibit slightly inward and outward motions of Zn (Ga) and O (N) at the surfaces, which has also been reported in previous studies [36,37]. The optimized supercell sizes ($a$, $b$) (unit in Å) are (3.300, 5.311) and (3.256, 5.291) for the pure ZnO and GaN $(1\bar{1}00)$ nanofilms, while (5.322, 5.689) and (5.289, 5.636) are observed for the pure ZnO and GaN $(11\bar{2}0)$ ones. For the same type, the pure ZnO and GaN nanofilms present almost the same structural parameters along the *a* or *b* axis, displaying negligible lattice mismatches. Thereby,

ZnO/GaN heterostructured nanofilms could be constructed within the same structural frame, as shown in Fig. 1. Similar to the pure ZnO and GaN nnofilms, the heterostructured nanofilms also display weak surface reconstructions with the surface Zn (Ga) and O (N) atoms possessing little inward and outward motions because of metallic interactions between Zn-Zn (Ga-Ga).

For further examining the structural stability of ZnO/GaN heterostructured nanofilms, the formation energy has been calculated according to the following equation [38,39]

$$E_f = [E_{tot} - (mE_{ZnO} + nE_{GaN})]/(m+n),$$

where $E_{ZnO}$ and $E_{GaN}$ are the potential energies (unit in eV/pair) of the wurtzite ZnO and GaN bulks, respectively. $E_{tot}$ is the total potential energy of the studied nanofilms. $m$ and $n$ are the numbers of Zn-O and Ga-N pairs in the ZnO/GaN heterostructured nanofilms. According to this definition, positive formation energy can be obtained, and a much smaller value indicates that nanofilm exhibits much better stability. As shown in Fig. 2, the formation energy increases integrally with the increasing GaN ratio for both $(1\bar{1}00)$ and $(11\bar{2}0)$ heterostructured nanofilms. In the former, it increases monotonically, whereas it shows an oscillating increase in the latter. Because of the positive $E_f$, Zn (Ga) and O (N) atoms in both the pure ZnO and GaN films prefer to $sp^3$ hybridization rather than $sp^2$ one. Besides, the pure ZnO nanofilms are more stable than the GaN ones for the same type, which is strongly dependent on the surface Zn-O and Ga-N pairs. In other words, the surface Ga-N pairs are less stable than the surface Zn-O ones. Due to the same numbers of the less stable surface Ga-N pairs, the same type of heterostructured nanofilms exhibit similar structural stability in the range of GaN ratio from 0.2 to 0.8.

*3.2 Electronic properties*

Fig. 3 shows electronic band structures of $(1\bar{1}00)$ and $(11\bar{2}0)$ ZnO/GaN heterostructured nanofilms as well as those of pure ZnO and GaN films. For the $(1\bar{1}00)$ nanofilms, the pure ZnO exhibits a direct band gap of 3.000 eV at Γ point (see Fig. 3a), whereas the pure GaN displays an indirect band gap of 3.158 eV (see Fig. 3c). For the $(11\bar{2}0)$ nanofilms, both the pure ZnO and GaN are direct band gap semiconductors with the gaps of 3.393 and 3.318 eV at Γ point (see Figs. 3d and 3f). These band gaps are beyond the energy range of visible light, suggesting that the pure ZnO and GaN nanofilms cannot absorb the visible light. However, for ZnO/GaN heterostructured nanofilms, their band gaps can be in the energy range of solar spectrum. Figs. 3b and 3e show the band structures of $(1\bar{1}00)$ type with GaN ratio of 0.375 and $(11\bar{2}0)$ with the GaN ratio of 0.3, respectively. In view of these figures, the heterostructured nanofilm with $(11\bar{2}0)$ frame exhibits a direct band gap, whereas the $(1\bar{1}00)$ nanofilm is indirect. But, owing to the small energy difference $\Delta E$ depicted in the inset of Fig. 3c, the pure GaN $(1\bar{1}00)$ nanofilm can be regarded as a quasi-direct band gap semiconductor. Such the quasi-direct band gap feature can also be found in the $(1\bar{1}00)$ ZnO/GaN heterostructured nanofilms. The band gaps are 2.469 and 2.668 eV for the $(1\bar{1}00)$ (GaN ratio of 0.375) and $(11\bar{2}0)$ (GaN ratio of 0.3) heterostructured nanofilms, respectively, which are much smaller than those of the pure ZnO and GaN nanofilms. Such narrow band gaps in ZnO/GaN heterostructured nanofilms are mainly attributed to the effects of type-II band offset [13,14]. Besides, the GaN ratio might have significant influences on the band structures of heterostructured nanofilms.

In order to clarify the effects of GaN ratio, the band gaps of ZnO/GaN heterostructured $(1\bar{1}00)$ and $(11\bar{2}0)$ nanofilms as a function of the GaN ratio are investigated and shown in Fig. 4. Of particular interest is that the band gaps of both $(1\bar{1}00)$ and $(11\bar{2}0)$ nanofilms decrease first and then increase as the GaN ratio increases, showing flexible tunability of band gaps. When the GaN ratio is 0.375, a minimum band gap of 2.469 eV is observed in the $(1\bar{1}00)$ heterostructured nanofilms. For the $(11\bar{2}0)$ nanofilms, the band gap of 2.668 eV is minimal as the GaN ratio reaches to 0.3. These band gaps are in the energy range of visible light spectrum and correspond to the wavelength of 500 and 460 nm or so. Furthermore, in the range of GaN ratio 0.1 to 0.75, all heterostructured nanofilms exhibit narrow band gaps, indicating optical absorption in the visible light spectrum.

### 3.3 Spatial charge distribution and optical properties

As is well known, an efficient photocatalyst not only possesses a suitable band gap, but also needs an effective spatial separation of photo-generated electron-hole pairs. Thus, the charge distributions of VBM and CBM are great significant for understanding the potential applications in photocatalytic field. Fig. 5 displays the charge distributions of $(1\bar{1}00)$ and $(11\bar{2}0)$ heterostructured nanofilms with the GaN ratios of 0.375 and 0.3, where the nanofilms exhibit minimal band gaps. For the $(1\bar{1}00)$ nanofilm, the VBM is mainly dominated by the GaN region and composed of N-$p$ state with little part from O-$p$ near the interface, while the CBM is mainly contributed by O-$p$ orbital of the ZnO region with little contribution from Zn-s orbital and N-$p$ state. Similar results have also been observed in the $(11\bar{2}0)$ nanofilm, suggesting that the interface has weak influence on the carrier species. In other words, the charge distributions of the VBMs (holes) and CBMs (electrons) are localized and come from different regions. Previous study has verified that the interfaces of heterostructured semiconductors can generate an effective potential barrier to block electrons and holes combining [40]. And thus, it might be expected that the ZnO/GaN heterostructured nanofilms can not only possess effective spatial separation of carriers, but also have low recombination of electron and hole pairs.

Due to the remarkable tunable band gaps that cover a wide range of the solar spectrum, ZnO/GaN heterostructured nanofilms could hold great promise for visible light harvesting. For exploring that, the optical properties are investigated by computing the complex dielectric function. Subsequently, the absorption coefficient $I(\omega)$ can be obtained according to the following equation [41]

$$I(\omega) = \sqrt{2}\omega[\sqrt{\varepsilon_1(\omega)^2 + \varepsilon_2(\omega)^2} - \varepsilon_1(\omega)],$$

where $\varepsilon_1(\omega)$ and $\varepsilon_2(\omega)$ are the real and imaginary parts of the dielectric function, and $\omega$ is a given frequency. The spectra of absorption coefficient for both the $(1\bar{1}00)$ and $(11\bar{2}0)$ ZnO/GaN heterostructured nanofilms with various GaN ratios are shown in Fig. 6. By comparasion with the pure ZnO and GaN films, the heterostructured nanofilms exhibit superior optical properties in the visible light spectrum. Such enhanced visible light absorption in the heterostructured nanofilms is mainly attributed to the narrow band gaps. For instance, $(1\bar{1}00)$ heterostructured nanofilm with GaN ratio of 0.375 displays obvious photo-absorption in the range of solar spectrum 400 to 560 nm due to its small band gap of 2.469 eV. For $(11\bar{2}0)$ heterostructured nanofilm with GaN ratio of 0.3, an absorption spectral range of 400 to 460 nm is presented owing to the band gap of 2.668 eV. Perhaps, the distinct absorption spectra in the $(1\bar{1}00)$ and $(11\bar{2}0)$ ZnO/GaN heterostructured nanofilms could be attributed to the different GaN ratios and interfaces.

For the low-dimensional nanostructures, size effects on electronic properties become remarkable and cannot be neglected [38,42-44]. Consequently, this raises a question: what are the relationships between the optical properties and the sizes of ZnO/GaN heterostructured nanofilms? Considering that the optical properties are mainly dependent on the electronic band structures, we have calculated the band gaps of heterostructured nanofilms as a function of the thickness to explore this question qualitatively, where all the films have the same GaN ratio $x = 0.5$. As shown in Fig. 7a, the band gaps of both $(1\bar{1}00)$ and $(11\bar{2}0)$ nanofilms decrease as their thicknesses increase. Such a reversely relationship could imply that much thicker nanofilms would show much narrower band gaps. In addition, a much thicker nanofilm exhibits much better structural stability (see Fig. 7b). Thereby, it could be inferred that both the $(1\bar{1}00)$ and $(11\bar{2}0)$ ZnO/GaN heterostructured nanofilms with larger thicknesses exhibit better structural stabilities, and could also display more superior optical absorption of visible light because of their narrower band gaps, suggesting their great promise for the visible light harvesting.

## 4. Conclusions

In summary, we have investigated the structural stabilities, electronic properties, and optical absorption of $(1\bar{1}00)$ and $(11\bar{2}0)$ ZnO/GaN heterostructured nanofilms by the spin-polarized DFT calculations. The formation energy, band structures, spatial charge distribution, and optical properties are calculated to explore the effects of GaN ratio and nanofilm thickness. The major conclusions are summarized as follows,

(1) The structural stability of ZnO/GaN heterostructured nanofilms roughly rises as the GaN ratio decreases. In the range of GaN ratio ($0.3 \leq x \leq 0.8$), all heterostructured nanofilms exhibit fairly structural stability. Moreover, a much thicker nanofilm displays better structural stability.

(2) All ZnO/GaN heterostructured nanofilms exhibit semiconductor features with the band gaps strongly depended on the GaN ratio and the film thickness. For both the $(1\bar{1}00)$ and $(11\bar{2}0)$ nanofilms, their band gaps decrease first and then increase with the increasing GaN ratio, while decrease always as the thicknesses increase. Due to narrower band gaps, the heterostructured nanofilms display more superior optical absorption in the visible light spectrum than the corresponding pure ZnO and GaN ones.

(3) For both the $(1\bar{1}00)$ and $(11\bar{2}0)$ heterostructured nanofilms, charge distributions of the VBMs and CBMs are mainly dominated by GaN and ZnO regions with N-$p$ and O-$p$ states, respectively. Such spatial separation of carriers might give rise to low recombination of electron and hole pairs.

Our results of ZnO/GaN heterostructured nanofilms with good structural stabilities, narrower band gaps, spatial separation of electrons and holes, and superior optical properties could promote their potential applications for the solar energy harvesting and converting.

## Acknowledgements

This work is supported by the National Natural Science Foundation of China (Grant Nos. 11204232 and 11374237). The authors acknowledge the computional support from the HPC Platform, Xi'an Jiaotong University.

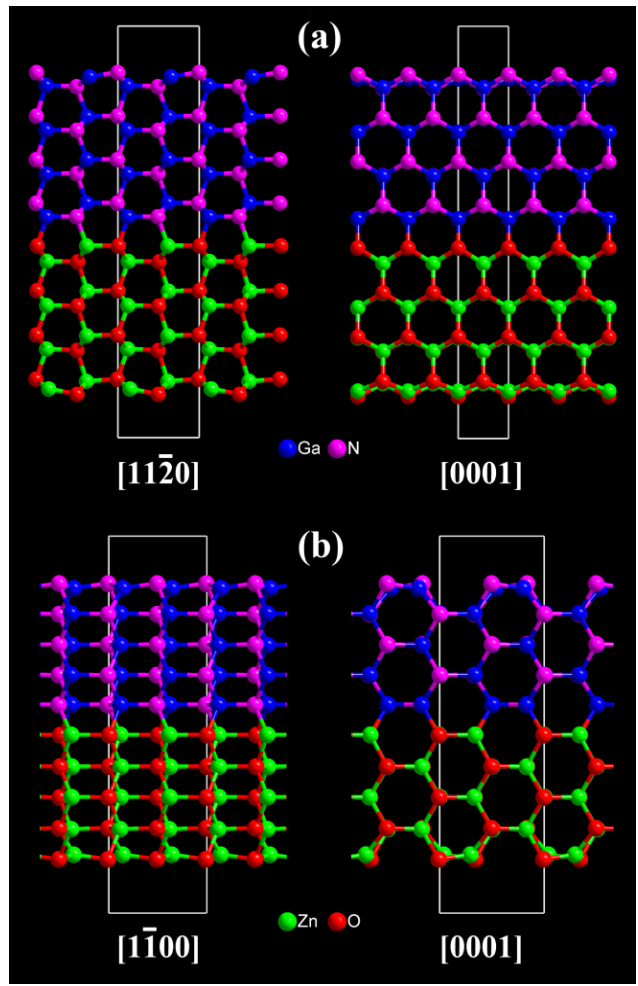

**Fig. 1.** Optimized structures of (a) (1$\bar{1}$00) and (b) (11$\bar{2}$0) ZnO/GaN heterostructured nanofilms with given 8 and 10 layers viewed from different directions. Solid rectangle denotes the supercell size. Blue, pink, green, and red balls stand for Ga, N, Zn, and O atoms, respectively.

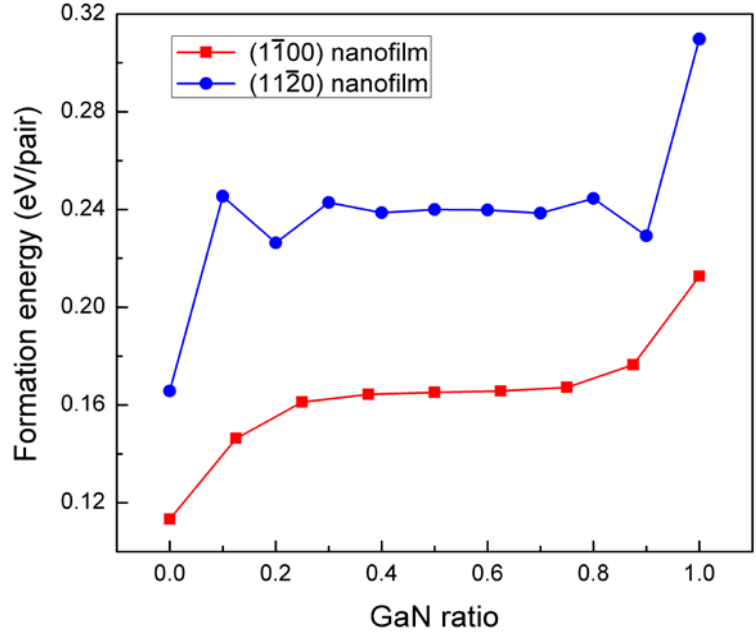

**Fig. 2.** Formation energy of ZnO/GaN heterostructured nanofilms as a function of GaN ratio. The GaN ratios of 0.0 and 1.0 represent the pure ZnO and GaN nanofilms, respectively.

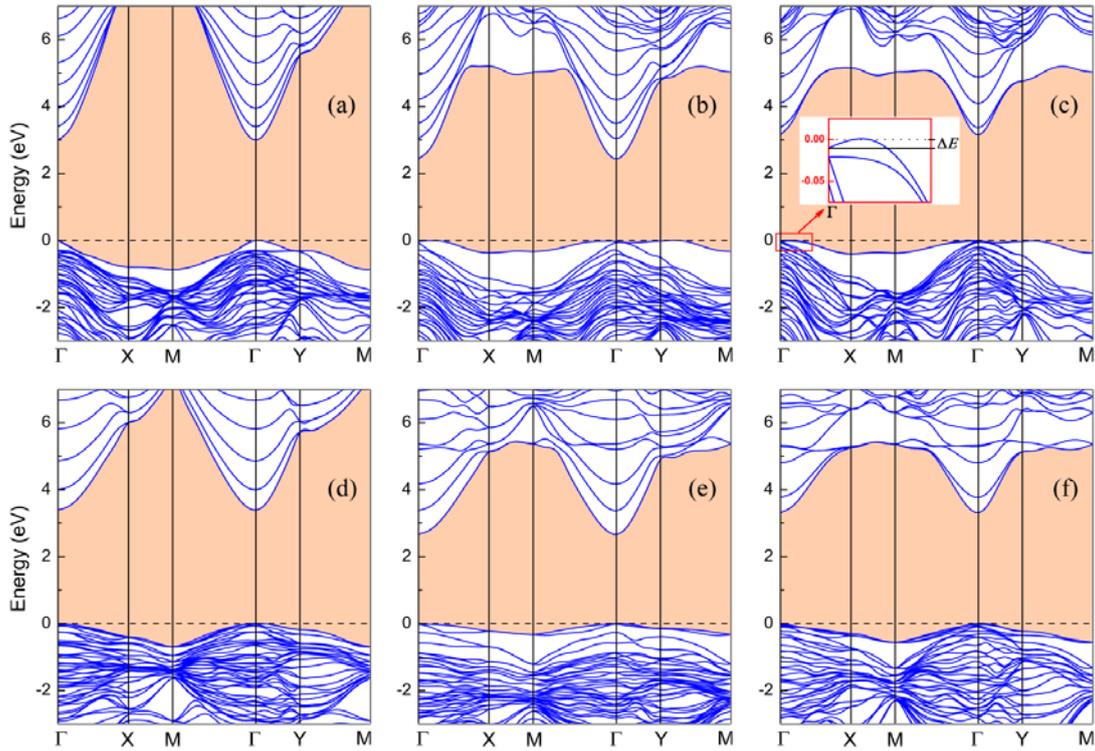

**Fig. 3.** Band structures of (a) pure ZnO, (b) ZnO/GaN heterostructure with GaN ratio of 0.375, and (c) pure GaN $(1\bar{1}00)$ nanofilms; (d) - (f) show pure ZnO, ZnO/GaN heterostructure with GaN ratio of 0.3, and pure GaN $(11\bar{2}0)$ nanofilms, respectively. The band structures are calculated by using HSE06. The energy of highest occupied level is set to zero and denoted by the dash line.

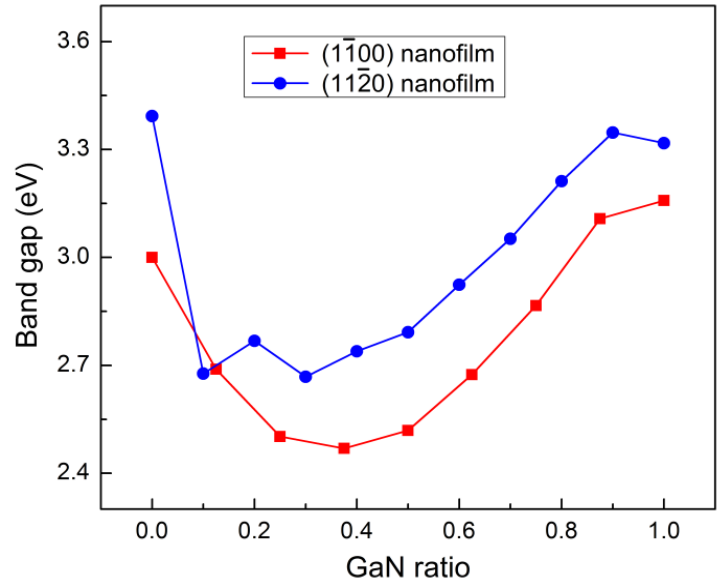

**Fig. 4.** Band gaps of ZnO/GaN heterostructured nanofilms as a function of GaN ratio. The ratios of 0 and 1 represent the pure ZnO and GaN nanofilms, respectively.

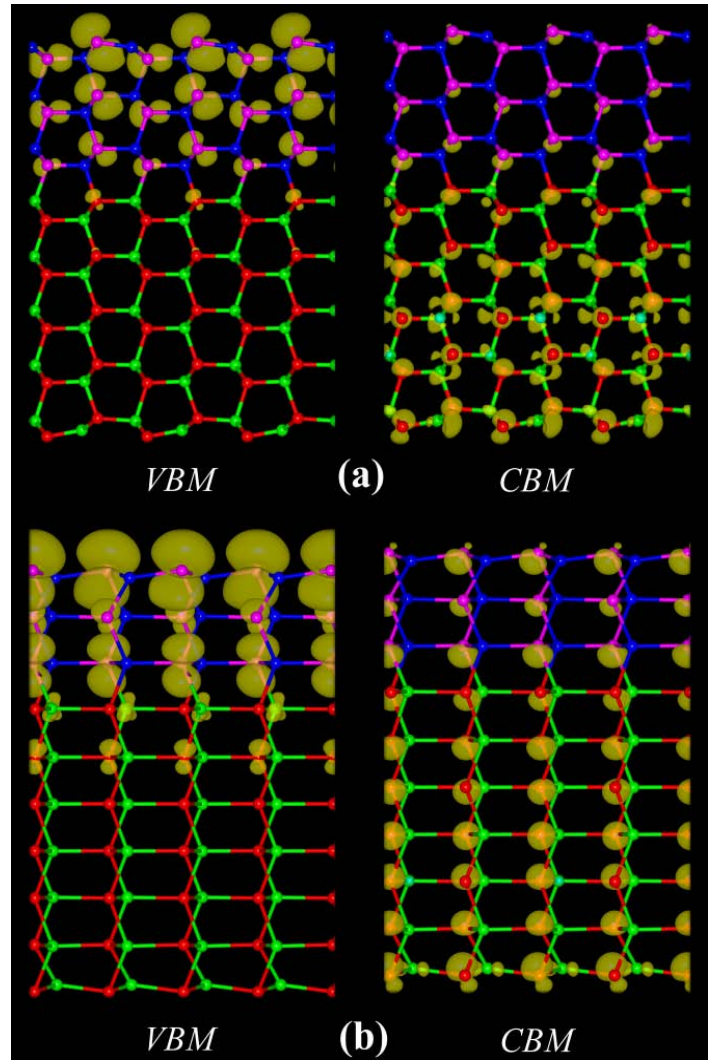

**Fig. 5.** Spatial charge distribution of VBMs (left panel) and CBMs (right panel) for (a) $(1\bar{1}00)$ and (b) $(11\bar{2}0)$ ZnO/GaN heterostructured nanofilms with GaN ratios of 0.375 and 0.3, respectively. The isovalue of charge density is set to the same value to clarify the differences of contributions. Green, red, blue, and pink balls represent Zn, O, Ga, and N atoms, respectively.

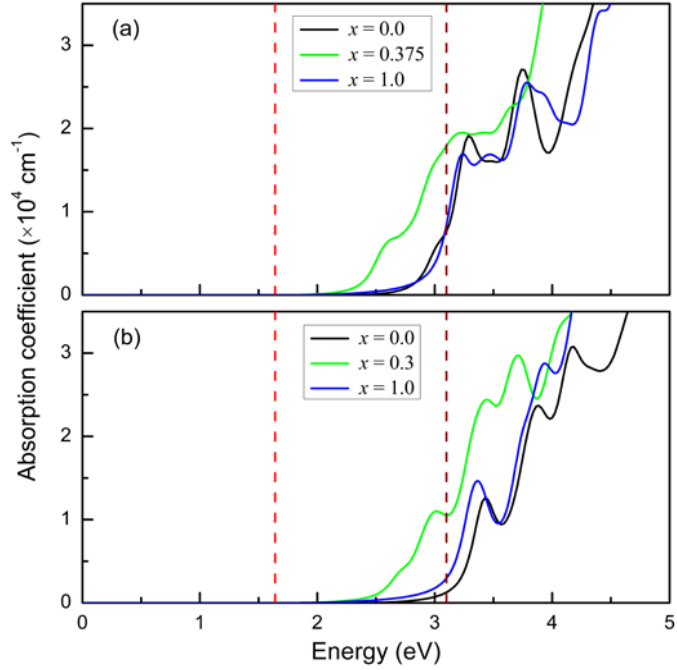

**Fig. 6.** Spectrums of absorption coefficient for (a) $(1\bar{1}00)$ and (b) $(11\bar{2}0)$ ZnO/GaN heterostructured nanofilms with different GaN ratios $x$. Vertical dash lines denote the energy range of visible light spectrum (i.e., 1.64 ~ 3.10 eV correspond to 760 ~ 400 nm).

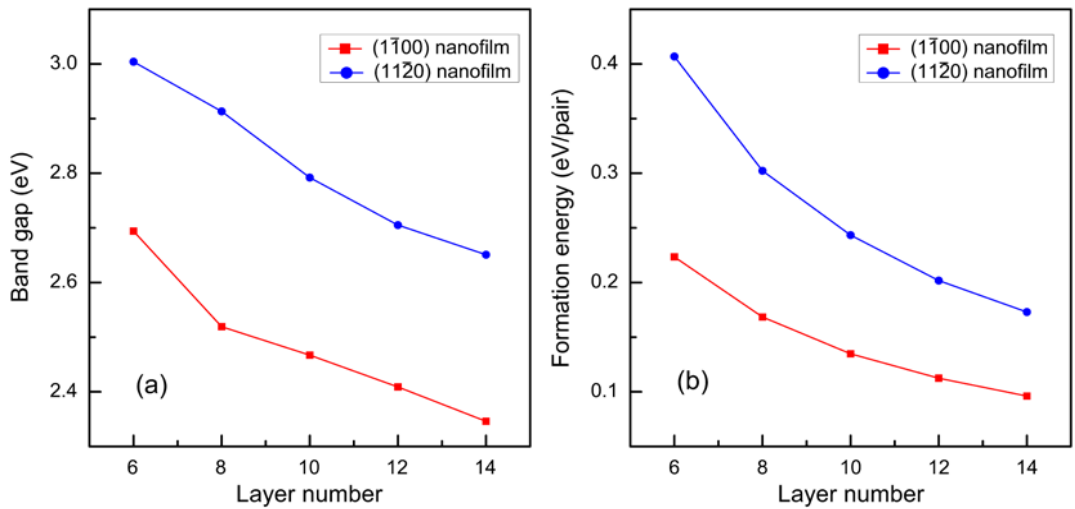

**Fig. 7.** Band gap (a) and formation energy (b) of ZnO/GaN heterostructured nanofilms as a function of the layer number (i.e., nanofilm thickness). Noted that all the nanofilms shown here have the same GaN ratio of 0.5.